\documentclass[prd,twocolumn,superscriptaddress]{revtex4-1}

\usepackage{graphicx, amsfonts, amsmath, amssymb, amscd,  theorem, exscale,
epic, eepic, epsfig}
\usepackage{xcolor}
\usepackage{orcidlink}

\usepackage{hyperref} %

\usepackage{makeidx}
\makeindex

\usepackage{xurl}

\newcounter{Figure}
\theoremstyle{plain}
\theoremheaderfont{\scshape}

\theorembodyfont{\upshape}

\theoremheaderfont{\scshape\small} \theorembodyfont{\scshape\small}

\newcommand{\real}{ {\mathbb R} }

\newcommand{\slap}{\mbox{$ \triangle \mkern -13mu / \ $}}
\newcommand{\nlap}{\mbox{$ \nabla \mkern -13mu / \ $}}
\newcommand{\dlap}{\mbox{$ div \mkern -13mu / \ $}}
\newcommand{\Dlap}{\mbox{$ D \mkern -13mu / \ $}}

\newcommand{\clapa}{\mbox{$ curl \mkern -23mu / \ $}}

\newcommand{\be}{\begin{equation}}
\newcommand{\ee}{\end{equation}}
\newcommand{\bea}{\begin{eqnarray}}
\newcommand{\eea}{\end{eqnarray}}
\newcommand{\beas}{\begin{eqnarray*}}
\newcommand{\eeas}{\end{eqnarray*}}

\newcommand{\Lie}{ {\mathcal L} }

\begin{document}

\title{Gravitational Wave Displacement and Velocity Memory Effects}

\author{Lydia Bieri \orcidlink{0000-0002-2469-3409}}
\email{lbieri@umich.edu}
\affiliation{Department of Mathematics, University of Michigan, Ann Arbor, MI 48109-1120, USA}

\author{Alexander Polnarev \orcidlink{0009-0009-8870-2982}}
\email{a.g.polnarev@qmul.ac.uk}
\affiliation{School of Physical and Chemical Sciences, Queen Mary University of London, London E1 4NS, United Kingdom}


\date{\today}


\begin{abstract} 
In this article, we compare in detail the linear
and nonlinear approach to the Gravitational Waves Displacement and Velocity Memory (GWDM and GWVM) effects. We consider astrophysical situations that give rise to gravitational waves with GWVM effect, i.e. with a residual velocity (the so-called ``velocity-coded memory") and discuss the possibility of future detection of the GWVM effect. 
\end{abstract} 

\maketitle

\tableofcontents

  \section{Introduction, Historic Overview}
\label{intro}

The main purpose of this article is two-fold. First, we present new results on an astrophysical scenario producing Gravitational Waves Velocity Memory (GWVM) in sections \ref{veldis}, \ref{nonla1} and in \ref{conclusions} Conclusions. 
The GWVM was first studied by L.P. Grishchuk and A.G. Polnarev in \cite{GrPo1}. 
Second, the remainder of the article presents a self-contained overview of the Gravitational Waves Displacement Memory (GWDM), in particular the contribution to this memory found by Ya.B. Zel'dovich and A.G. Polnarev in \cite{zeldovichpolnarev} as well as the contribution by D. Christodoulou in \cite{chrmemory}. The key point is to make this discussion accessible to a broad audience of physicists and mathematicians. Therefore, sections 
\ref{linear}, \ref{nonlaversuslinear} and \ref{nonlinear} of the main text explain the two contributions to the GWDM and their differences. The mathematical derivations of these results are provided in the appendix to make the paper self-contained. The current format allows readers to choose at what depth to make use of the article. They may either choose to concentrate on the main text with the physical explanations, or they may choose to also read the mathematical derivations provided in the appendix. Whereas the results in sections \ref{veldis}, \ref{nonla1}, \ref{conclusions} are new, the topics discussed in the remainder of this article (including the appendices) are well-known, and we give the precise citations for the latter.

Generally, massive accelerating objects produce gravitational waves. This includes the flyby of very dense objects, or the scenarios of black holes or neutron stars orbiting each other and finally merging. 
These waves are fluctuations of the curvature of the spacetime. They travel at the speed of light along the light cones (that is null hypersurfaces) of the spacetime. 
Gravitational waves travel a very long distance from their source to arrive at our detectors. During that time, 
the wave signals become extremely weak to be detected. However, 
detectors of LIGO (Laser Interferometer Gravitational-Wave Observatory) have overcome these challenges and have been able to 
detect gravitational waves since 2015, \cite{ligodetect1, ligodetect2, ligodetect3}. 
The first detected gravitational waves were generated by the merging of two black holes in binary black hole systems. 
In 2017, for the first time gravitational waves were measured from the merging of two neutron stars in a binary system by the LIGO Scientific Collaboration and Virgo Collaboration \cite{ligodetect4}. In the meantime, measurements of gravitational waves have become quite frequent.

Nearly half a century ago, in 1969, Joseph Weber announced that he detected gravitational waves with his aluminium bar detector with sensitivity approximately ten million times worse than the sensitivity of LIGO in 2015. Today it is absolutely clear that this was a false detection. However, the first decade after Weber's false detection, theoretical physicists tried to find mechanisms explaining the generation of gravitational waves of such high amplitude. 
See for example \cite{synchr}.

In 1974, in an attempt to demonstrate that such a detection was impossible from the theoretical point of view, Zel'dovich and Polnarev considered the system which seemed these days to be the most favorable for the detection of gravitational waves. Namely, a hypothetical super dense cluster consisting of compact objects (black holes or neutron stars) situated in the center of our Galaxy. Gravitational radiation generated by such a cluster was coursed by flybys of compact objects around each other. As a byproduct of the corresponding calculations it was shown that gravitational radiation bursts change the spacetime permanently \cite{zeldovichpolnarev}. 
In a detector this would show as a permanent residual displacement of the test masses after the wave burst has passed.  In \cite{braginskyg, braginsky} this was called the memory effect of gravitational waves. 

In this paper we will call this effect Gravitational Waves Displacement Memory (GWDM). The GWDM first calculated by Zel'dovich and Polnarev corresponds to weak field and small velocity approximation when it is possible to use the linearized theory of gravity, i.e. linear approximation. In this approximation the memory effect is proportional to 
an overall change of the second time derivative of the source's quadrupole moment \cite{zeldovichpolnarev}.

For sources of gravitational radiation like the
flyby of compact objects  or like the merger of
two black holes the test masses will go to rest after the passage of the gravitational wave burst (see above). 

Quite different a kind of gravitational waves bursts still in linear approximation was studied by Grishchuk and Polnarev \cite{GrPo1}. Imagine that after such bursts of gravitational radiation passed through two separated test masses they continue to  move relative to each other during an extended interval of time, considerably exceeding the duration of the burst. We will see below (section \ref{veldis}) that such bursts  could be generated in some astrophysical situation. 
We call this the Gravitational Waves Velocity Memory (GWVM) effect. 

Taking into account that the test masses move
along geodesics of the spacetime, their displacements (GWDM) or residual velocities (GWVM) visualizes the changes of the geodesics in space-time
and hence the spacetime itself.
Both GWDM and GWVM effects in linear approximation are discussed in section \ref{linear}.

In 1991, Christodoulou \cite{chrmemory} using the fully nonlinear theory of gravitation computed the gravitational wave memory effect for bound systems such as binary black holes, that is for strong gravitational fields and large velocities when linear approximation does not work. The Christodoulou result \cite{chrmemory} can be interpreted as consisting of two parts. One of them is due to the change of final mass and velocities in the source. 

We will call this the Ordinary  Contribution to the memory effect (OCME) because this is the analogue of the Zel'dovich and Polnarev memory effect but applied to a bound system with arbitrary strong gravitational fields and arbitrary high velocities. Another part of the Christodoulou memory effect is related to the total energy radiated away in form of gravitational waves which propagate along null geodesics. For this reason, we call this part the Null Contribution to the Memory Effect (NCME)  which is absolutely new in contrast to the ordinary memory effect.  
The details for the Christodoulou contributions to memory, OCME and NCME, is given in section \ref{nonlinear}. 

We emphasize that all above mentioned memory effects  (\cite{zeldovichpolnarev}, \cite{GrPo1} and \cite{chrmemory}) 
manifest themselves as permanent displacements or velocities of the test masses in a detector like LIGO (\cite{ligodetect1}) or LISA (\cite{Audley2017}) (while the detection of the gravitational waves background of low frequency by NANOGrav (\cite{nanograv1}) and similar projects is based on statistics of variations in arrival time of radio signals from numerous radio pulsars (\cite{nanograv1, pta2, pta3, pta4})). 

Since the pioneering days of memory research, this field has grown tremendously, and various new contributions to the gravitational wave memory have been found. Here, we give a short historic overview of the most important results that are most relevant to our work. Given the large and increasing literature on this topic, this list is by no means complete, and it would not be feasible to aim for the latter. 
The pioneering works \cite{zeldovichpolnarev} and \cite{chrmemory} were followed by several contributions \cite{blda1, blda2, braginsky, braginskyg, will, thorne, thorne2, jorg}. In recent years, the field has grown fast, shedding light on various aspects of memory 
\cite{1lpst1, 1lpst2, lbdg1, lbdg2, lbdg3, asht1, tolwal1, winicour, Lasky1, strominger, flanagan, flanagan2, favata, BGYmemcosmo1, bgsty1, twcosmo, lbatdgbw, winma2}. 
In particular, in \cite{1lpst1}, \cite{1lpst2} it was shown that in spacetimes that are solutions of the Einstein-Maxwell equations, the electromagnetic field contributes to the gravitational null memory.

L. Bieri showed \cite{lydia4} for the Einstein vacuum equations as well as the Einstein equations coupled to neutrinos a variety of new gravitational wave memory effects in asymptotically flat spacetimes, where the metric approaches the metric of Minkowski spacetime towards infinity at a slow rate. (See Appendix \ref{new1} for detail). 

Cosmological aspects of gravitational memory effects (including the role of red-shift and gravitational lensing)  were discussed  by L. Bieri, D. Garfinkle and N. Yunes in \cite{BGYmemcosmo1}. See also  A. Tolish and R.M. Wald \cite{twcosmo}, and L. Bieri, D. Garfinkle, S.-T. Yau \cite{bgsty1}. In particular, for the Einstein equations with a cosmological constant, as shown in \cite{bgsty1}, in de Sitter spacetime the null memory is enhanced by a factor including the redshift. Similarly, a corresponding result for FLRW (Friedmann-Lema\^itre-Robertson-Walker) spacetime was obtained in \cite{twcosmo}. Gravitational waves in the $\Lambda$CDM cosmological models were considered in \cite{BGYmemcosmo1}. It was shown there that gravitational wave memory is enhanced by a redshift-dependent factor. It was also shown in  \cite{BGYmemcosmo1} that this memory can be affected by gravitational lensing.

In \cite{Lasky1}, 
P. Lasky, E. Thrane, Y. Levin, J. Blackman and Y. Chen suggested a way to detect gravitational 
wave memory with LIGO by stacking events, i.e. to detect GWDM after numerous gravitational wave events rather than from one single gravitational wave event.   

Outside of General Relativity, 
Bieri and Garfinkle derived \cite{lbdg2} the analogues of both contributions (Zel'dovich-Polnarev and Christodoulou) to memory 
for the Maxwell equations. These were the first analogues outside General Relativity. 
The search for analogues of the memory effect in other theories, in particular in quantum field theories, has grown as well. See for instance \cite{strominger} or \cite{WaldTm1}. \\ 

In this article, we compare in detail the linear and nonlinear approach to gravitational waves memory and discuss some new astrophysical aspects of the GWVM effect. 

This paper is organized as follows. Sections \ref{linear} and \ref{veldis} contain the linear analysis of (GWDM) and (GWVM) correspondingly. Then  sections \ref{nonlinear} and \ref{nonla1} are devoted to the nonlinear aspects of (GWDM) and (GWVM). 
Finally, in section \ref{conclusions} we present some ideas for possible future detection of GWDM  and GWVM effects. 
In particular, new results on the GWVM are obtained in sections \ref{veldis}, \ref{nonla1} and in \ref{conclusions}. Namely, section \ref{veldis} yields new insights using linear approximation, whereas section \ref{nonla1} treats the same scenario via nonlinear analysis deriving new nonlinear contributions. To compute the latter, previous mathematical results from \cite{lydia3} are used. In order to keep this paper self-contained, we cite the most important mathematical derivations in the appendix.

Another goal of this and forthcoming papers is the application of the rigorous mathematical results from the above cited works to those problems of gravitational wave astronomy, where gravitational fields are strong and  velocities are high.

Starting with \cite{sta} many rigorous mathematical results have been obtained studying the Einstein equations in the fully nonlinear theory (see above). The nonlinear analysis in our paper is based on the results obtained in \cite{lydia3}.

\section{GWDM in Linear Approximation}
\label{linear} 

 In this section, we present the Zel'dovich-Polnarev contribution to GWDM \cite{zeldovichpolnarev}. 
 
Two test masses in free fall with initial separation $d^i$  have a relative acceleration (see \cite{lanlif}, \cite{MTW} for example) given by 
\begin{equation} \label{geodev1gen}
{\frac {d^2 \Delta x^j} {d t^2} = - R^j_{0i0}d^i}
\end{equation}

with $R^j_{0i0}$ denoting corresponding components of the Riemann curvature tensor. We use Latin letters  for the spatial coordinates and $0$ for time-coordinate. 
In linear approximation at a large distance $r$ from the source 
\begin{equation} \label{approxQR1}
{R_{0i0j}} = -{\frac {1} r} P \left [ {\frac {{d^4}{Q_{ij}}} {d {t^4}}} \right ]
\end{equation} 
Here $Q_{ij}$ is the quadrupole moment of the source given by
\begin{equation} \label{quadru1}
{Q^{ij}} = \int {d^3} x {T^{00}} {x^i}{x^j}
\end{equation} 
and $P[]$ denotes ``projected 
to be orthogonal to the radial direction and trace free.''  
It follows that (a) a residual velocity would be proportional
to the difference between ${d^3}{Q_{ij}}/d{t^3}$ at large positive time and large negative time; 
and (b) a permanent displacement would be proportional
to the difference between ${d^2}{Q_{ij}}/d{t^2}$ at large positive time and large negative time. 
However, regarding (a), we note that for 
large positive or negative time the widely separated gravitating objects (generating gravitational waves described by the Riemann tensor above) should move without accelerations, i.e. with constant 
velocities. Hence 
\begin{equation}
{\frac {d^2} {d {t^2}}} {Q^{ij}} = {\sum _k} {m_k}{v _k ^i}{v_k ^j}
\label{d2q}
\end{equation}  
where the sum is over all objects with the $k$th object having mass $m_k$ and velocity ${\vec v}_k$.  Then it follows
from equation (\ref{d2q}) that at large positive and negative times ${d^3}{Q_{ij}}/d{t^3}$ vanishes, and therefore
there is no residual velocity. On the other hand, regarding (b), there is a permanent displacement given by 
\begin{equation}     \label{zpm1}
\Delta {x^j} = \frac {d_i}{r}  P \left[  \sum _k [(m_k
v _k ^i v_k ^j)_{\infty} - (m_k v _k ^i v_k ^j)_{-\infty}]  \right]
 \end{equation}
This (\ref{zpm1}) is precisely the GWDM in linear approximation found in \cite{zeldovichpolnarev}.

\section{GWVM in Linear Approximation}
\label{veldis}

This section is devoted to the further development of the idea of velocity-coded memory, formulated in \cite{GrPo1}, and to a more detailed application of this idea to measuring the effects of gravitational memory in the future.
Let us (following to \cite{GrPo1}) consider the following astrophysical situation:

A supermassive black hole of mass $M$ is surrounded by a large accretion disk of mass $m_d =q_d \cdot M$, where $q_d < 1$, and of radius $r_d$.  

A less massive black hole of mass $m=q_m \cdot M$, where $q_m<1$, moves perpendicular to the plane of the disk and intersects it at radius $r < r_d$.

In this paper we assume that the central supermassive black hole is Schwarzschild, i.e. a non-rotating black hole. 

We denote the direction of motion of the smaller black hole when it is very close to the point of intersection with the disk as the $x$-direction. The intersection point corresponds to $x=0$.

 We can consider the gravitational field of the disk as uniform if $\sqrt {r_{int}^2+x^2}< r_d$.

After crossing the disk the black hole of mass $m$ experiences a jump of acceleration which is equal to 
\be
\Delta a_m = G \sigma_d 
\ee
where 
\be
\sigma = \frac{m_d}{4 \pi r_d^2}
\ee
is the average surface density of the disk. 
Hence it is 
\be
\Delta a_m = \frac{G m_d}{4 \pi r_d^2} = \frac{q_d r_g c^2}{8 \pi r_d^2}
\ee
where $r_g$ is the gravitational radius of the central black hole. 

The jump of the acceleration results in the jump of the third derivative of the $xx$-component of the quadrupole moment. The jump is 
\be
\Delta \frac{d^3 D_{xx}}{dt^3} = m \Delta \frac{d^3 x^2}{dt^3} = 
\frac{q_m r_g c^2}{2 G} 2 \Delta a_m V_m 
\ee
with $V_m$ being the velocity of the less massive black hole at the moment when it crosses the disk. 
Obviously, it is 
\be
V_m \approx 
\sqrt{\frac{2GM}{r_d}} = c\sqrt\frac{r_g }{r_d} \ . 
\ee

Let us consider now a remote detector consisting of two test masses separated by a distance $l$. 
The jump in the third derivative of the quadrupole moment generates the jump in the $x$-component of the relative velocity of these two test masses: 
\be \label{**vxDxx}
\Delta v_x = \Delta \dot{h}_{xx} \approx \frac{2G}{c^4 L} \Delta \frac{d^3 D_{xx}}{dt^3} 
\ee
where $L$ is the distance between the detector and the system of the two black holes. 
Thus, we have 
\be \label{*vx}
\Delta v_x \approx \frac{c}{2 \pi} q_m q_d \frac{r_g^{\frac{5}{2}} l }{r_d^2 r^{\frac{1}{2}} L }
\ee
The separation between the test masses will be changing with this relative velocity during the time interval $\delta t$ of order (see (10))
\be
\delta t \approx \frac{r_d}{V_m} \approx \frac{r^{3/2}_d}{c r_g^{1/2}} \approx 10^{-5}s\frac{r_d^{3/2}}{r_g^{3/2}} \frac{M}{M_{\odot}}  \ . 
\ee
If, for example, $r_d=100 r_g$ and $M=3\cdot 10^9 M_{\odot}$, $\delta t \approx 3\cdot 10^7 s \approx 1$ year. 
As a result the distance between the test masses will be 
\be \label{*xx}
\Delta x \approx \Delta v \delta t \approx \frac{q_m q_d r_g^2 l}{2 \pi r_d L} \ . 
\ee

That is, for a very long time the velocity of the test masses will be non-zero, and the displacement memory will keep building up. 
We point out the important difference between this result and the memory studied in \cite{zeldovichpolnarev} and \cite{chrmemory} and explained in the previous sections. Namely, that in the latter cases 
the relative velocity of the test masses will be non-zero for a very short time only during the burst, 
that could be a fraction of a second for the cases studied there. After that short time, the test masses will return to rest. 
Whereas equations (\ref{**vxDxx})-(\ref{*xx}) show that the velocity of the test 
masses will be non-zero for a very long time. 

Thus, we not only estimated the jump in relative velocity between two test masses in the detector, as was done in \cite{GrPo1}, but also estimated the resulting displacement between these masses and the characteristic time for which this displacement is gained.

This is the velocity-coded memory, GWVM. 
 Even though this may look like a weak signal, it is integrated over a long interval of time, making the effect larger and 
increasing the chances of a future detection. 
For very rough estimates, i.e. as a first approximation, 
we used the linear approximation. However, one can imagine such a set of parameters of the system considered above that the nonlinear contribution may be significant. 

\section{Nonlinear Analysis versus Linear Approximation } 
\label{nonlaversuslinear}

Let us start from the fundamental Einstein equations \cite{MTW}, \cite{lanlif} describing the generation and propagation of gravitational fields. Setting the constants $G=c=1$, these equations read 
\be \label{ET}
R_{\mu \nu} - \frac{1}{2} R g_{\mu \nu}  \ = \ 8 \pi \ T_{\mu \nu} 
\ee
$\mu, \nu = 0,1,2,3$ correspond to four spacetime dimensions. 
(Generally in this paper, Greek letters $\alpha, \beta, \gamma, \cdots$ denote spacetime indices.) 
$R_{\mu \nu}$ denotes the Ricci curvature tensor, $R$ is the scalar curvature, 
$g_{\mu \nu}$ is the metric tensor and $T_{\mu \nu}$ is the stress-energy tensor. 
The non-gravitational fields represented by stress-energy on the right hand side of (\ref{ET}) have to satisfy their own equations, and one has to solve the corresponding coupled system. 

In vacuum, i.e. in the case of pure gravity, the right hand side of (\ref{ET}) is zero and the equations reduce to 
the Einstein vacuum (EV) equations 
\be \label{EV*}
R_{\mu \nu} = 0  \ . 
\ee 

Recall that the black hole solutions of Schwarzschild and Kerr are solutions of the Einstein vacuum equations (\ref{EV*}).

Recall also that the Riemann curvature tensor $R_{\alpha \beta \gamma \delta}$ is related to the Weyl tensor, $W_{\alpha \beta \gamma \delta}$, which is trace-less, by the following formula: 
\bea
R_{\alpha \beta \gamma \delta} & = & W_{\alpha \beta \gamma \delta}  \nonumber \\ 
& & 
  \hspace{-1cm}  + 
\frac{1}{2}  ( g_{\alpha \gamma}  R_{\beta \delta }  +  g_{\beta \delta } R_{\alpha \gamma}  -  
g_{\beta \gamma} R_{\alpha \delta }  -  g_{\alpha \delta} R_{\beta \gamma} ) \nonumber \\ 
& & 
  \hspace{-1cm}  - \frac{1}{6} ( g_{\alpha \gamma} g_{\beta \delta}  - g_{\alpha \delta} g_{\beta \gamma}  ) R  \ .   \label{RiemWeyl}
\eea

Obviously, in the vacuum case (\ref{EV*}) $R_{\alpha \beta \gamma \delta}  =  W_{\alpha \beta \gamma \delta}$. 
 
A. Einstein found gravitational waves 
as solution for the linearized field equations in his 1916 and 1918 papers \cite{ae1916*2}, \cite{ae1918*1}. However, 
the first proof that gravitational waves exist in the fully nonlinear theory was given by Y. Choquet-Bruhat \cite{ychb1} in 1952 in her proof of the local existence and uniqueness of solutions to the EV equations \ref{EV*}. 
Denote by 
$\Box$ the d'Alembertian.  
Writing the Ricci curvature in equation (\ref{EV*}) in terms of the metric 
and using harmonic coordinates (see \cite{Wald1}) (the latter are often called wave coordinates because the coordinates satisfy a wave equation) 
and omitting indices in the arguments on the right hand side, we obtain 
\be    \label{EVwave1}
\Box g_{\alpha \beta} = N_{\alpha \beta} (g, \partial g)   
\ee 
with $N_{\alpha \beta} (g, \partial g)$ denoting nonlinear terms which are quadratics with respect to $\partial_{\gamma} g_{\mu \nu}$. 
Thus, we have a system of wave equations with these nonlinearities on the right hand side. 

The gravitational wave memory is computed from the geodesic deviation equation, where the acceleration of nearby geodesics is given by the Riemann curvature tensor \cite{zeldovichpolnarev}, \cite{chrmemory}. Recall from above that, when considering the Einstein vacuum equations (\ref{EV*}), the Riemann curvature is equal to the Weyl curvature. The permanent displacement of the geodesics is then computed by integrating twice the Weyl tensor. 
The Weyl tensor obeys its own set of equations.  
As we see later the Bianchi identities \cite{ONeill} play a crucial role when we derive the memory. 
Later on, we shall write these equations in a frame that is better adapted to study gravitational radiation. 

We now turn to the well-known Bianchi identities \cite{ONeill}. 
In the EV case the Bianchi identities are reduced to the following simple equation 
\be  \label{Bianchi2}
D^{\alpha} W_{\alpha \beta \gamma \delta} \ = \ 0 \ . 
\ee

Often it is helpful to make use of the ``analogy" between general relativity (GR) and electromagnetism (EM). 
Therefore, we decompose the Weyl curvature tensor $W_{\alpha \beta \gamma \delta}$ into the components defined as follows: 
\begin{eqnarray}
{E_{ab}} := {W_{a0b0}} \label{Wel1}
\\
{H_{ab}} := {\textstyle {1 \over 2}} {{\epsilon ^{ef}}_a}{W_{efb0}} \label{Wma1}
\end{eqnarray}
where we denote by $t$ the time and refer to the time-coordinate as the $0$-coordinate, we use Latin letters $a, b, c, \cdots$ for the spatial parts, and 
where $\epsilon _{abc}$ is the spatial volume element (see for instance \cite{Wald1}), related to the spacetime volume element by 
${\epsilon_{abc}} = {\epsilon_{0abc}}$. 
We then call $E_{ab}$ the electric and $H_{ab}$ the magnetic parts of the Weyl tensor $W_{\alpha \beta \gamma \delta}$. (See for instance \cite{sta}.) 
Even though the Maxwell equations are linear and the Einstein equations nonlinear, thinking of analogies in the behavior of the electric and magnetic fields in EM and the electric and magnetic parts of the Weyl curvature in GR has been fruitful. 

Writing the Bianchi identities for the Weyl curvature components (\ref{Wel1})-(\ref{Wma1}), we find again a useful analogy to the Maxwell equations. Later on, we shall introduce further decomposition of the Weyl curvature tensor and these equations that will be best adapted to studying gravitational waves and the memory effect. Moreover, this treatment lays open more analogies between GR and EM. 

Astrophysical processes like the merger of black holes are described by the initial value problem for the Einstein vacuum equations (\ref{EV*}). 
 
 The set of initial conditions in General Relativity for (\ref{EV*}) consists of $(H_0, \bar{g}_{ij}, k_{ij})$ with $i, j = 1, 2, 3$, where $H_0$ is the initial space-like hypersurface, $\bar{g}_{ij}$ the initial metric, namely a $3$-dimensional 
Riemann metric, and $k_{ij}$ a $2$-covariant symmetric tensor field on $H_0$ (the second fundamental form) \cite{ONeill}, 
satisfying 
the constraint equations \cite{Wald1}: 
\beas
\overline{\nabla}^i k_{ij} - \partial_j  k_i^i & = &  0 \\ 
\overline{R} - |k|^2 + (tr k)^2 & = &  0
\eeas
where 
the overbar denotes quantities in the initial spacelike hypersurface $(H_0, \bar{g}_{ij})$, that is 
$\overline{\nabla}$ is 
the covariant derivative and $\overline{R}$ is the scalar curvature of $(H_0, \bar{g}_{ij})$. 
We will mainly be interested in asymptotically flat systems, that is systems where 
there exists a coordinate system $(x^1, x^2, x^3)$ in a neighborhood of infinity such that with 
$r = (\sum_{i=1}^{3} (x^i)^2 )^{\frac{1}{2}} \to \infty$, it is, 
\bea
\bar{g}_{ij} \ & = & \ \delta_{ij} \ + \ h_{ij} \  + \ o \ (r^{- \frac{3}{2}})  \label{imetric33} \\ 
k_{ij} \ & = & \ o (r^{-\frac{5}{2}})  \label{iff33}
\eea
with $h_{ij} = O(mass/r)$. These systems were investigated in \cite{lydia3}. 
See the latter for details.

Linear analysis used in previous sections \ref{veldis} and \ref{linear} 
is not valid 
for strong fields or high velocities.
Instead the latter require 
to deal with  the fully nonlinear Einstein equations. A good example 
is a bound system of black holes, where the bodies spiral around each other, lose energy and finally merge. 

Another difference between the method used in the previous two sections and the method to be presented in the next two sections is the following: The quadrupole moment approximation (\ref{approxQR1}) is not valid. 

The nonlinear method to be discussed next can be used to solve (\ref{ET}) and 
(\ref{EV*}). 
This is extremely important when we use the geodesic deviation equation for two test masses representing the detector of gravitational waves. As follows from (\ref{EV*}) and (\ref{RiemWeyl}) we can replace the Riemann tensor by the Weyl tensor: 
\be \label{geodW}
\frac{d^2 x^j}{dt^2} \ = \ - W^j_{ \ 0i0} x^i \ . 
\ee

\section{Nonlinear Analysis of GWDM}
\label{nonlinear}

The purpose of this section is to investigate in detail what goes into the right hand side of equation (\ref{geodev1gen}), 
and to derive all contributions (null and ordinary) to the gravitational wave memory. 
In order to do so, in particular to have a full description of the dynamics at $\mathcal{I^+}$ we will have to look first at the relevant equations near the source (see appendix \ref{detailsextra}) and then compute them 
at future null infinity $\mathcal{I^+}$. 
Thereby we shall find a term causing the ordinary contribution to memory and another term that produces the null contribution to memory. 
More precisely, we shall show that 
the ordinary contribution results from a change of a particular component of the Weyl tensor over retarded time, whereas the null contribution is due to the integral over all of retarded time of the energy density called the Bondi news. The latter is denoted by $| \Xi  |^2$. In other words, the null contribution to memory is caused by the waves themselves radiating away energy. 
We write for the energy $F$ 
\be \label{recallF1}
F = \frac{1}{2} \int_{- \infty}^{+ \infty} | \Xi  |^2 du
\ee
where {$F/4 \pi$ is the total energy radiated away (to infinity) per unit 
solid angle in a given direction \cite{chrmemory}. 

$ | \Xi  |^2$ denotes the Bondi news and $u$ is the retarded time. This $F$ is a function on $S^2$ and it is the main ingredient in the final formula for the null contribution. 
The precise formula for the null contribution to the memory is given by (\ref{*C1*}).

The formula for the full memory effect should be of the form 
\be \label{memfirst}
\Delta x = - \frac{d}{r} \big( {\small \mbox{``ordinary"} + \mbox{``null"}} \big)  \ , 
\ee
where ``ordinary" is the ordinary contribution and ``null" the null contribution to memory involving (\ref{recallF1}). 
The ordinary contribution in (\ref{memfirst}) has a similar structure like (\ref{zpm1}) but is not an approximation rather an exact expression. 
In the case of a binary coalescence it is computed from the final rest mass of the body and its final (recoil) velocity. 
The null contribution in (\ref{memfirst}), as is the right hand side, is also an exact expression and is completely different from the  former. It involves $F$ from (\ref{recallF1}), in which the energy density $ | \Xi  |^2$ is integrated over retarded time $u$, and this $F$ will be integrated over the sphere $S^2$  expressing the fact that we compute the permanent displacement in the plane orthogonal to the direction of observation. 
Thus, there are the following three key facts about the null contribution to memory that make it a truly new effect: 
\begin{itemize}
\item[(1)] It is sourced by the quadratic term $ | \Xi  |^2$ describing the energy density. 
\item[(2)] This term $ | \Xi  |^2$, which originally may be small, is integrated over a long interval of retarded time $u$. Thereby the resulting quantity $F$ from (\ref{recallF1}) becomes large. Recall that $F/4 \pi$ is the total energy radiated away (to infinity) per unit solid angle in a given direction, and we can think of this as the 
gravitational waves themselves generate additional gravitational fields. 

Working with equation (\ref{geodev1gen}) and the Einstein equations (\ref{ET}), it was shown that that stress-energy tensors like an 
electro-magnetic field tensor \cite{1lpst1}, \cite{1lpst2} or neutrino radiation tensor \cite{lbdg1} also contribute to the null gravitational memory. 

In future papers we shall work with equation (\ref{geodev1gen})  and  Einstein equations (\ref{ET}) to show that these stress-energy tensors also contribute to the null gravitational memory in the astrophysical scenario presented in sections \ref{linear} and \ref{nonla1}.

\item[(3)] Finally, this energy from (\ref{recallF1}) is integrated over the sphere, that is $F(\xi')$ is integrated over $\xi' \in S^2$, meaning that 
a contribution of this energy $F$ from (\ref{recallF1}) 
is added from each point on the sphere. 
This is new and exclusive for the null memory, being in stark contrast to the ordinary memory which is not integrated over the sphere. 
Rather the ordinary memory, in a binary coalescence, is computed from the final rest mass of the body and its final (recoil) velocity, where no integration is involved. 
\end{itemize}

It is remarkable that the null memory involves two integrals, namely the integral over retarded time $u$ in (\ref{recallF1}) and the integral of $F$ over the sphere. 
The null contribution to the permanent displacement is a cumulative effect, building up during the time that the gravitational wave burst 
passes through the detector. See formula (\ref{allmemgen1}) or with all the details formula (\ref{*C1*all}). We derive the latter in appendix \ref{detailsextra}.

In order to compute (\ref{memfirst}) in detail, we have to introduce new quantities and convenient foliations of the spacetime. 
We use the same spacetime foliations as in \cite{lydia1}, \cite{lydia2}, \cite{sta}.  
The time $t$ foliates the spacetime into spacelike hypersurfaces $H_t$. 
The second foliation of the spacetime is given by a function $u$ (retarded time), namely a foliation into outgoing light cones, that is outgoing null hypersurfaces $C_u$. 
In order to calculate (\ref{recallF1}) and (\ref{memfirst}), we need to take intersections of the spacelike hypersurfaces $H_t$ with the light cones $C_u$. This goes as follows: 
Let $S_{t,u} = H_t \cap C_u$ be the $2$-surfaces of intersection. The latter are topological spheres, more precisely they are diffeomorphic to the sphere $S^2$ and play the role of wave fronts. 
We shall call $\theta_1, \theta_2$ given on $S_{t,u}$ the spherical variables. 
Further, it is $r = r(t,u) = \sqrt{\frac{Area(S_{t,u})}{4 \pi}}$. For the details of this construction, we refer to \cite{sta}.

We recall that 
future null infinity $\mathcal{I}^+$ is defined to be the endpoints of all future-directed null geodesics along which $r \to \infty$. 
It has the topology of $\real \times \mathbb{S}^2$. 
Thus a null hypersurface $C_u$ intersects $\mathcal{I}^+$ at infinity in a $2$-sphere $S_{\infty, u}$. (See for instance \cite{Wald1}.) 

In the following, all operators on $S^2$ are denoted with a slash, $\dlap$, $\clapa \ $, $\nlap$, $\slap$, and are automatically operators with respect to the standard metric on $S^2$. 

In the following, when computing the ordinary and null memory for a binary black hole coalescence, we shall make use of the rigorous mathematical results for such systems, \cite{chrmemory}, \cite{lydia3}. We also give more details on these spacetimes and the derivation of both the ordinary and null memory in the appendix (\ref{detailsextra}).

We introduce a more suitable frame to study gravitational waves, namely, the following null frame taking into account the structure of the light cones. (See \cite{chrmemory}.) 
We may think of a detector like LIGO, where $m_0$ denotes the reference test mass, and $m_1$, $m_2$ two test masses. 
The use of the geodesic deviation equation is standard. Here, we use the notation given in \cite{chrmemory}. 

Choose the vectors $\vec{E}_1$ and $\vec{E}_2$ 
in the directions of the masses $m_1$ and $m_2$ initially. 
Then the $(x^1, x^2)$ plane is horizontal like the mirrors and lasers in LIGO. 
Denote by $x^j_{(A)}$ the coordinates of the test mass $m_A$ with $A = \{ 1, 2 \}$. 
That is, the index in brackets refers to the $A^{th}$ test mass $m_A$. 
Capital Latin letters $A, B, C, \cdots$ take values in $\{ 1, 2 \}$. 
Assume that the source is in the $\vec{E}_3$ direction. Then the null normals $\vec{L}$ and $\underline{\vec{L}}$ are $\vec{L} = \vec{T} + \vec{E}_3$ and $\underline{\vec{L}} = \vec{T} - \vec{E}_3$, where $\vec{T}$ is the unit timelike vectorfield as given in \cite{chrmemory}. This gives the null frame $\{ \vec{e}_1, \vec{e}_2, \vec{e}_3 =\underline{\vec{L}},  \vec{e}_4 = \vec{L}  \}$. 
In this new frame, the leading order curvature component in (\ref{geodW}) is $W_{A 3 B 3}$. 
This curvature component 
has a limit at future null infinity $\mathcal{I}^+$ and we call this limit  $\underline{A}_{AB}$ (see \cite{chrmemory}, \cite{lydia4}): 
\be \label{defcurva1}
\underline{A}_{AB} = \lim_{ t \to \infty} r W_{A 3 B 3|_{C_u,}} 
\ee
The motion of the test masses in the $\vec{E}_3$ direction is negligible \cite{chrmemory} and we have for the motion in the plane the equation 
\be \label{gA1}
\ddot{x}_{A(C)} = - \frac{1}{4} r^{-1} \underline{A}_{AB} x^B_{(C)} + O(r^{-2}) \ . 
\ee
Here, $\ddot{x}_{A(C)} = \frac{d^2 x_{A(C)} }{dt^2}$ denotes the $A$-component of the acceleration vector of the 
test mass $m_C$, and $\dot{x}_{A(C)}$ shall denote the corresponding velocity component. 
As the test masses are at rest at their positions $x^A_{(A)}$ before the arrival of the gravitational wave burst, we are given 
the initial conditions 
$x^i_{(A)} (+ \infty) = d_0 \delta^i_A$, $\dot{x}^i_{(A)} (+ \infty) = 0$, and ignoring lower order terms, we obtain 
\be \label{gA2}
\ddot{x}_{A(B)}= - \frac{d_0 n^D_{(B)}}{4r} \underline{A}_{AD} 
\ee
with $n^D_{(B)}$ the components of the corresponding unit vector. 
Equation (\ref{gA2}) is just the rewritten equation (\ref{geodW}) but in a more suitable frame and with initial conditions taken into account. 
Integrating (\ref{gA2}) gives 
\be \label{dotxintA1}
\dot{x}_{A(B)} (u)  = - \frac{1}{4} d_0 r^{-1} \int^{+ \infty}_u n^D_{(B)} \underline{A}_{AD} (u') du' 
\ee
From the mathematical results \cite{chrmemory}, \cite{lydia4} for systems that we are investigating here, we know that  the following equations hold at future null infinity $\mathcal{I^+}$, and 
we also derive these equations in the appendix \ref{future}: 
\begin{eqnarray}
\frac{\partial \Xi_{AB} }{\partial u} &=&-\frac{1}{4} \underline{A}_{AB}   \label{1XiAT1} \\ 
\frac{\partial \Sigma_{AB} }{\partial u} &=& - \Xi_{AB} \label{1SigmaT1} 
\end{eqnarray}
where $\Sigma_{AB}$ is the limit of the shear of the outgoing light cone, and $\Xi_{AB}$ the limit of the shear of the incoming light cone. 
That is, the radiative amplitude per unit solid angle for one of these shear quantities is given by 
$\Xi_{AB}$, while that for the other shear quantity is $\Sigma_{AB}$. 
Recall that 
$|\Xi|^2$ appears in (\ref{recallF1}) for $F$ which is the total energy radiated away per unit angle in a given direction. 
See appendix \ref{future} for further details. 
Next, using (\ref{1XiAT1}) in (\ref{gA1}) yields 
\be \label{vel1}
\dot{x}_{A(B)} (u)  = - d_0 r^{-1}  n^D_{(B)}  \Xi_{AD} (u) 
\ee 
Integrating again and using (\ref{1SigmaT1}) yields 
\bea
x_{A(B)} (u) - x_{A(B)} (+ \infty) & = &  \nonumber \\ 
 - d_0 r^{-1}   n^D_{(B)} (\Sigma_{AD} (u) - \Sigma_{AD} (+ \infty))   \label{idispl}
\eea
Then we obtain 
\be
\triangle x_{A(B)} = - \frac{d_0}{r} n^D_{(B)} \big( \Sigma_{AD} (- \infty) - \Sigma_{AD} (+ \infty)  \big)  
\ee
which can be written as  
\be \label{pdispl*}
\triangle x_{A(B)} = - \frac{d_0}{r} \big( \Sigma_{AB} (- \infty) - \Sigma_{AB} (+ \infty)  \big)  
\ee 
This equation describes the GWDM in the nonlinear theory including both the ordinary and the null contributions to this memory \cite{chrmemory}. 
The quantity $ \big( \Sigma_{AB} (- \infty) - \Sigma_{AB} (+ \infty)  \big)  $ on the right hand side of equation (\ref{pdispl*}) is computed as the solution of a system of equations at future null infinity $\mathcal{I}^+$: 
\bea
\dlap \dlap (\Sigma^- - \Sigma^+) 
 \ & = & \ (P^- - P^+) 
- 2 F \nonumber \\ 
\label{Pall27} \\ 
\clapa \ \ \dlap (\Sigma^- - \Sigma^+)  \label{Qlimits*1*} \ & = & \   Q^- - Q^+ \ = \ 0  \nonumber \\ 
 \label{Qall27} 
\eea 
where $P$ and $Q$ are limits of different curvature components (see appendix \ref{future} for details), and $\Sigma^+ = \lim_{u \to \infty} \Sigma(u)$, correspondingly for the other limits with upper $+$ respectively upper $-$. 
We derive these equations in appendix \ref{detailsextra}. 

In the scenario where two bodies are coming in at non-relativistic speed, spiral around each other and finally merge to one body with final rest mass $M^*$ and with final (recoil) velocity $\vec{V}$ relative to the initial center-of-mass frame, we know from 
(\ref{rhoA2}), (\ref{sigmaA2}) and the corresponding limits in section \ref{future} that 
$Q = O(|u|^{- \frac{1}{2}})$. Therefore, the right hand side of (\ref{Qall27}) is zero. 
On the other hand, from sections \ref{future} and \ref{Cauchy} we know that the terms on the right hand side of (\ref{Pall27}) do not vanish. See also \cite{chrmemory} and \cite{lydia4} for further details. 

In the following, $X, Y, V, \xi$ denote vectors, but we shall drop the arrows to simplify the notation.

The solution 
$(\Sigma^- - \Sigma^+)(X, Y)$ of (\ref{Pall27}) and (\ref{Qall27}) for this scenario of the binary coalescence is computed at an arbitrary pair $X, Y$ of vectors in $\real^3$ tangent to $S^2$, at $\xi$. Here, $\xi \in S^2 \subset \real^3$ is the direction of observation. 
Thus, the plane spanned by $(X, Y)$ is orthogonal to $\xi$.

\bea
& & \hspace{-0.5cm} \small{  (\Sigma^- - \Sigma^+)(X, Y)  = }   \nonumber \\ 
& & \hspace{-0.5cm} \small{ - \frac{2 M^*}{( 1 - |V|^{2} )^{\frac{1}{2}}} \cdot f (X, Y, V, \xi)  } \nonumber \\ 
& &  \hspace{-0.5cm} \small{ - \frac{1}{2 \pi} \int_{|\xi'| = 1} 
\large(  F - F_{[1]} \large) (\xi') } 
\small{ \cdot g (X, Y, \xi , \xi' )
d S^2 (\xi') }   \label{allmemgen1} \nonumber \\ 
\eea 
Here, subscript $[1]$ denotes the projection onto the sum of the $0^{th}$ $(l=0)$ and $1^{st}$ $(l=1)$ eigenspaces of $\slap$, that is on the sum of the $(l=0)$ and $(l=1)$ spherical harmonics. 
On the right hand side of (\ref{allmemgen1}), in the first term, $f (X, Y, V, \xi)$ gives the projection of $V$ onto the $X, Y$ plane, thus the plane transverse to the observation, and ensures the trace-free property, similarly in the second term, $g (X, Y, \xi , \xi' )$ gives the projection of $\xi'$ onto the $X, Y$ plane, thus the plane transverse to the observation, and ensures the trace-free property. The solution computed in (\ref{allmemgen1}) has the properties that it is symmetric, trace-free and transverse, the $X, Y$ plane being orthogonal to the direction of observation $\xi$. 
The exact formulas that were derived by Christodoulou in \cite{chrmemory} with explicit $f (X, Y, V, \xi)$ and $g (X, Y, \xi , \xi' )$ 
are given in (\ref{*C1*all}) respectively (\ref{*P*}) and (\ref{*C1*}) in the appendix \ref{Christodouloumemory}. 
Using $(\Sigma^- - \Sigma^+)(X, Y)$ in equation (\ref{pdispl*}), the first term on the right hand side of (\ref{allmemgen1})  describes the ordinary contribution to memory and the second term on the right hand side of (\ref{allmemgen1}) describes the null contribution to memory. 
That is, the former corresponds to the contribution to memory derived by Zel'dovich and Polnarev \cite{zeldovichpolnarev} but applied to binary coalescence, the latter is the contribution to memory derived by Christodoulou \cite{chrmemory}.

\section{Nonlinear Analysis of GWVM}
\label{nonla1}

In this section we apply the nonlinear analysis to 
the scenario from section \ref{veldis}}.

Instead of going in retarded time $u$ all the way to $+ \infty$ respectively $- \infty$, we can concentrate on a finite but large enough interval 
$[u^-, u^+]$. Thus, here, ``upper indices" $+$ and $-$ mean the finite limits of $u$. 

The acceleration jump is seen as a jump in curvature, which happens in a very short time interval. At the detector, this burst 
arrives and lasts for the short time $\triangle u$. 
After this short time $\triangle u$, the velocity of the test masses stays constant over a longer time interval $\delta u$.

In order to compute this memory, we use some notation introduced in the appendix \ref{mainstructuresequations} and used already in section \ref{nonlinear} for a different situation. In particular we write for the leading order curvature component $W_{A 3 B 3} = \underline{\alpha}_{AB}$. 

This curvature component has the following limit at future null infinity $\mathcal{I}^+$: 
\be \label{defcurva2}
\underline{A}_{AB} = \lim_{ t \to \infty} r W_{A 3 B 3|_{C_u}} 
\ee
That is, like in the limit (\ref{defcurva1}) above, the curvature term $W_{A 3 B 3}$ attains its corresponding limit, namely (\ref{defcurva2}), in the new scenario. However, $\underline{A}_{AB}$ behaves very differently here in comparison to the performance of the corresponding quantity in section \ref{nonlinear}. 

Moreover, as in section \ref{nonlinear}, $\Sigma_{AB}$ denotes the limit of the shear for the outgoing light cone, 
$\Xi_{AB}$ the limit of the shear for the incoming light cone, and $| \Xi |^2$ is the energy density. 
However, these quantities behave differently in the current scenario. The equations for these quantities are derived in the appendix \ref{detailsextra}. In order to use them in the present section, we take into account their particular behavior in $u$.

We collect a few facts. 
For the spacetimes considered here it follows from (\ref{uchihat}) that (\ref{1XiAT1}) holds, 
it follows from (\ref{chihat}) that (\ref{1SigmaT1}) holds, 
and from (\ref{codazzi2}) that (\ref{Ldchibund1}) holds. 
As the small black hole is moving along the $x$-direction, as described in section \ref{veldis}, and the acceleration jump occurs in the $x$-direction, we obtain the following. 
The acceleration jump $\triangle a$ gives a jump in the curvature component $\underline{\alpha}_{xx}$:

\be \label{Auf**}
\triangle \underline{\alpha}_{xx} = \triangle a
\ee
This gives the limit

\be \label{Aa**}
\triangle \underline{A}_{xx} =  \lim_{t \to \infty} r \triangle \underline{\alpha}_{xx|_{C_u,}} 
\ee

Let us denote by $y^i_{(B)}$ the coordinates of test masses $m_{(B)}$  with $B = \{ 1, 2 \}$. 
Then the geodesic deviation equation reads

\be \label{gA22}
\ddot{y}_{A(B)} = - \frac{1}{4} r^{-1}  \underline{A}_{AC} y^C_{(B)}  \ . 
\ee 

Taking into account the initial conditions, it is: 
\[
\ddot{y}_{A(B)} = - \frac{d_0}{4r} n^D_{(B)} \underline{A}_{AD} 
\]
Then we can write 
\bea
 \triangle \ddot{y}_{A(B)} 
= - \frac{d_0}{4r} n^D_{(B)} \triangle \underline{A}_{AD} & & \label{ya1} \\ 
 \triangle \dot{y}_{A(B)} 
  = - \frac{d_0}{r} n^D_{(B)} \triangle \Xi_{AD}  \nonumber \\ 
= - \frac{d_0}{r} n^D_{(B)} \triangle \underline{A}_{AD} \triangle u & &\label{ya2} \\ 
\triangle y_{A(B)}   =  - \frac{d_0}{r} n^D_{(B)} \triangle \Sigma_{AD} & & \nonumber \\ 
= \frac{d_0}{r} n^D_{(B)} \triangle \underline{A}_{AD} \frac{(\triangle u)^2}{2} + \nonumber \\ 
 \frac{d_0}{r} n^D_{(B)} \triangle \underline{A}_{AD} \triangle u \delta u  &&  \label{ya4}
\eea
where the equations are describing acceleration, velocity, and permanent displacement of the test masses, respectively. In the current scenario, we can simply compute $\triangle  \underline{A}_{AB}$ via $\triangle a$ from the data introduced in the description of the problem in section \ref{veldis}. Then these equations yield the memory of velocity (equation (\ref{ya2})) as well as the memory of displacement (equation (\ref{ya4})). 

Let us then compare (\ref{ya2}) and {\ref{ya4}} to the results of sections \ref{veldis} and \ref{nonlinear}. 
The main difference to the scenarios studied in section \ref{nonlinear} is that in the current scenario some of the quantities do not fall off with $u$ for a long time. 

Comparing the two terms on the 
the right hand side of equation (\ref{ya4}) (the first term is the null contribution while the second one is the ordinary contribution) and taking into account that 
$\triangle u$ is considerably smaller than $\delta u$ we can see that the null contribution to the GWDM in this scenario is considerably smaller than the ordinary contribution.

Comparing the ordinary contribution (the last term in (\ref{ya2})) 
to the result from section \ref{veldis} obtained by linear approximation 
we see that in the current scenario the quadrupole approach gives a good approximation for the ordinary contribution and therefore for the GWVM. It will be interesting to investigate other astrophysical scenarios where the parameters are different and the quadrupole approximation does not work anymore, and where the null contribution to the GWVM may be large. The authors plan to tackle these questions in further articles.

Finally, we can ask what happens if we wait for an even longer time. Then in the above scenario 
$\Xi_{AB}$ will start falling off in retarded time $u$. Therefore, the velocity of the test masses will go to zero, meaning that the velocity-coded memory will disappear. On the other hand the displacement memory will remain in the limit as a permanent displacement.

\section{Conclusions}
\label{conclusions}

We have seen that a small black hole moving in the direction perpendicular to the plane of the accretion disk around a more massive black hole generates a gravitational wave burst with a velocity-coded memory.

In section \ref{veldis}, using the quadrupole moment approximation, we compute the resulting displacement between the test  masses of a gravitational wave detector and the characteristic time for which this displacement is accumulated. In section \ref{nonla1}, we explore the same scenario with nonlinear analysis. A comparison of the results obtained in section \ref{veldis} and \ref{nonla1} shows that 
the quadrupole approach gives a good approximation for the ordinary memory for a reasonable set of astrophysical parameters. Nevertheless, it is possible to imagine such sets of parameters where the nonlinear effects are large and the quadrupole approximation does not work. The authors plan to investigate these questions in future work.

In a gravitational wave detector, so far, the time of observation has been of the order of the time of the burst itself. However, 
in order to detect gravitational waves of the type considered in section \ref{veldis} of
this article, we suggest to use the gravitational memory, which means that the duration of measurements is considerably longer than the duration of the burst of gravitational radiation. The increase of time of measurement gives a real chance to detect such signals.

We compare in this paper nonlinear and linear approaches. We emphasize that there are two types of contributions to gravitational wave memory, namely the ordinary and the null. The ordinary contribution is similar to the effect obtained by linear approximation but applicable to the case of  
strong fields and 
high velocities. In contrast to ordinary, the null contribution to gravitational wave memory is completely different. 
The null contribution is related with the fact that the energy of gravitational waves itself generates additional gravitational waves. 

In conclusion, we emphasize our hope that the future of gravitational wave memory (linear or nonlinear, ordinary or null) is bright.  \\ \\

\section*{Acknowledgments}

We thank the organizers of the workshop ``Gravitational Memory Effects: From Theory to Observation" held at the Queen Mary University of London in June 2023. In particular we thank Pau Figueras, David Nichols, Ali Seraj, and Juan A. Valiente Kroon for their organization and for making possible this volume. 
LB thanks the NSF, acknowledging support from the NSF Grant DMS-2204182.

\appendix

\section{Detailed Derivation of Null Memory}
\label{detailsextra}

\subsection{Main Structures and Equations}
\label{mainstructuresequations}

We consider the Einstein vacuum (EV) equations (\ref{EV*}) from above.

Let us denote by $(M,g)$ the spacetimes solving (\ref{EV*}). 
Astrophysical processes like the merger of binary black holes are described by solutions of these equations. More precisely, by the initial value problem for (\ref{EV*}). Hence, a realistic treatment of these processes requires solving the initial value problem for data describing physical processes. 
Therefore, we shall describe classes of initial data describing physical sources of gravitational radiation, solve the Einstein equations for this data, and thereby produce solution spacetimes from which the radiation field and memory can be read off. 
Precisely this information will lay open the new structures for the memory. 
These results on the dynamics in the radiation field (at future null infinity $\mathcal{I^+}$) are rigorous and emerge from the fully nonlinear treatment.

Recall from section \ref{nonlinear} the foliations of the spacetime by $t$ and $u$. 
We also recall from above that, 
regarding the $t$-foliation of the $4$-dimensional spacetime manifold, $T$ denotes the future-directed unit normal to $H_t$. Now, denote by $N$ the outward unit normal to $S_{t,u}$ in $H_t$. More precisely, it is $N = a^{-1} \frac{\partial}{\partial u}$ with lapse $a = | \nabla u |^{-1}$.  
We introduce the null frame $e_1, e_2, e_3, e_4$, with $\{ e_A \}, A = 1,2$ being a local frame field for $S_{t,u}$, and $e_3 = \underline{L}, e_4 = L$  a null pair. That is $g(e_4, e_3) = -2$. 
Note that the outgoing null vector field $e_4 = T+N$, the incoming null vector field $e_3 = T-N$. Further, we introduce $\tau_- := \sqrt{1 + u^2}$. And we use $k_{ij}$ for the second fundamental form of $H_t$.

$D$ or $\nabla$ are used to denote the covariant differentiation on $M$, whereas $\overline{\nabla}$ or $\nabla$ are used on the spacelike hypersurface $H$. Furthermore, we write operators on the surfaces $S_{t,u}$ with a slash. 
For a $p$-covariant tensor field $t$ tangent to $S$, $\Dlap_4t$ and $\Dlap_3t$ are the projections to $S$ of the Lie derivatives 
$\Lie_4 t$, respectively $\Lie_3 t$. \\

\begin{center}

\makebox{ \hspace{-1.8cm} \includegraphics[scale=0.7]{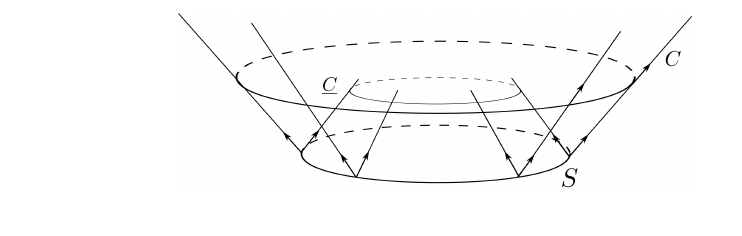}}

\end{center}

The figure shows an outgoing light cone $C$ and an ingoing light cone $\underline{C}$ from a spacelike $2$-surface $S$. $C$ is generated via the null vector fields $e_4$, and $\underline{C}$ is generated via the null vector fields $e_3$. 

The spacetime curvature at $S_{t,u}$ decomposes as follows: 
For any vectors $X, Y$ tangent to $S_{t,u}$ at a point and $\epsilon$ the area $2$-form of $S_{t,u}$, define 
$\alpha (X,Y) : = R(X, e_4, Y, e_4)$, $\underline{\alpha} (X,Y) : = R(X, e_3, Y, e_3)$, 
$2 \beta (X) : =  R(X, e_4, e_3, e_4)$, $2 \underline{\beta} (X, Y) : =  R(X, e_3, e_3, e_4)$, 
$4 \rho : = R(e_4, e_3, e_4, e_3)$, $2 \sigma \epsilon (X, Y) : = R(X, Y, e_3, e_4)$. 
Thus, the spacetime curvature at $S_{t,u}$ decomposes into the following components: the symmetric $2$-covariant tensorfields 
$\alpha$, $\underline{\alpha}$, the $1$-forms $\beta$, $\underline{\beta}$, and the scalar functions $\rho$, $\sigma$. 
For a full description of the curvature decomposition, see \cite{sta} or \cite{lydia2}. 

{\itshape Shears $\hat{\chi}(X,Y)$, $\underline{\hat{\chi}}(X,Y)$ and curvature component $\underline{\alpha}(X,Y)$: }
In the following, $l.o.t.$ stands for ``lower order terms" meaning ``more fall-off". 
Quantities crucial to describe gravitational radiation are the shears of the light cones, namely $\hat{\chi}$, $\underline{\hat{\chi}}$ being the traceless parts of the second fundamental forms as follows:  
$\chi (X, Y) := g(\nabla_{X} e_4, Y)$ is the second fundamental form computed for $S$ in the outgoing light cone $C_u$, 
respectively $\underline{\chi} (X, Y) := g(\nabla_{X} e_3, Y)$ is the second fundamental form computed for $S$ in the corresponding incoming light cone, 
for $X, Y$ tangent to $S_{t,u}$. 
The shears $\hat{\chi}$, $\underline{\hat{\chi}}$ are 2-tensors tangential to the surfaces $S_{t,u}$, and we denote their components by $\hat{\chi}_{AB}$, respectively $\underline{\hat{\chi}}_{AB}$. 
The traces $tr \chi$ and $tr \underline{\chi}$ of the second fundamental forms $\chi$ respectively $\underline{\chi}$ are the expansion scalars measuring how the area element of the surface $S$ changes along the outgoing respectively incoming light cone. 
The shear $\underline{\hat{\chi}}$ measures outgoing radiation, and is therefore a key player in understanding memory. 
Further, we introduce the torsion-$1$-form $\zeta$ to be defined by 
$\zeta_A := \frac{1}{2} g(D_{e_A} e_4, e_3)$. 
The shears obey the following well-known geometric equations \cite{ONeill}; 
the Codazzi equations (\ref{codazzi1})-(\ref{codazzi2}). 
\bea
\dlap \hat{\chi} & = &   - \hat{\chi} \cdot \zeta + \frac{1}{2} (\nlap tr \chi + \zeta tr \chi) - \beta  \nonumber \\ 
 \label{codazzi1} \\ 
\dlap \hat{\underline{\chi}} & = &  \hat{\underline{\chi}} \cdot \zeta + \frac{1}{2} (\nlap tr \underline{\chi} - \zeta tr \underline{\chi}) + \underline{\beta}   \nonumber \\  \label{codazzi2} \\ 
& = & \underline{\beta} + l.o.t.   \nonumber 
\eea

\subsection{Cauchy Problem for the Einstein Equations with Data Describing Physical Scenarios}
\label{Cauchy}

In this section, we present mathematically rigorous results for spacetimes describing the evolution of physical systems in GR.

In \cite{lydia3}, Bieri 
considered asymptotically flat initial data $(H_0, \bar{g}_{ij}, k_{ij})$ with $i, j = 1, 2, 3$, where $\bar{g}$ and $k$ are sufficiently smooth and  
for which there exists a coordinate system $(x^1, x^2, x^3)$ in a neighborhood of infinity such that with 
$r = (\sum_{i=1}^{3} (x^i)^2 )^{\frac{1}{2}} \to \infty$, it is, referring to this type of initial data and the corresponding spacetimes as (A):  
\bea
\bar{g}_{ij} \ & = & \ \delta_{ij} \ + \ h_{ij} \  + \ o_3 \ (r^{- \frac{3}{2}})  \label{initialdg1}  \\ 
k_{ij} \ & = & \ o_2 (r^{-\frac{5}{2}})  \label{initialdk1}   
\eea
with $h_{ij}$ being homogeneous of degree $-1$. 
In particular, $h$ may include a non-isotropic mass term. 

In \cite{sta}, D. Christodoulou and S. Klainerman proved the global nonlinear stability of Minkowski space using initial data of the following type, as $r \to \infty$, referring to this type of initial data and the corresponding spacetimes as (CK): 
\bea
\bar{g}_{ij} \ & = & \ (1 \ + \ \frac{2M}{r}) \ \delta_{ij} \ + \ o_4 \ (r^{- \frac{3}{2}}) \label{safg33} \\ 
k_{ij} \ & = & \  o_3 \ (r^{- \frac{5}{2}}) \ ,  \label{safk33}
\eea
where $M$ denotes the mass and is constant. 

In \cite{lydia1}, \cite{lydia2}, Bieri proved the global nonlinear stability for 
asymptotically flat initial data such that as $r \to \infty$, referring to this type of initial data and the corresponding spacetimes as (B): 
\bea
\bar{g}_{ij} \ & = & \ \delta_{ij} \ + \ 
o_3 \ (r^{- \frac{1}{2}}) \label{afgeng}  \\
k_{ij} \ & = & \ o_2 \ (r^{- \frac{3}{2}})    \ .    \label{afgenk}  
\eea
 
These stability proofs provide global existence and uniqueness results for small initial data, that is the data is supposed to be small enough in weighted Sobolev norms. Then global solutions to the Einstein equations are constructed that are asymptotically flat and causally geodesically complete. Thus, they do not develop any singularities. In addition, these proofs provide detailed information on the dynamics of the spacetimes, the behavior of their geometry and physical quantities, in particular this includes the behavior and fall-off of the curvature components and the shear towards future null infinity $\mathcal{I^+}$. Whereas the smallness assumption was needed to establish the existence of these solutions, the behavior along null hypersurfaces towards future null infinity of the spacetime is basically independent from the smallness. Therefore, even for large data, including black holes, we find that in the corresponding spacetimes the asymptotic results still hold for a portion of null infinity. See \cite{chrmemory}, \cite{lydia4}, \cite{lydia14}. Consequently, these rigorous descriptions of the behavior of the gravitational field at null infinity lay open information about physical scenarios including large data such as black holes. From these we can derive new results on gravitational radiation and memory for various such sources.

In the above works, it was shown that in the global future development of the initial data, the following terms have a decay behavior at infinity that is given by the following, 
for (A) spacetimes 
\bea
\underline{\alpha} \ & = & \ O \ ( r^{- 1} \ \tau_-^{- \frac{5}{2}}) \label{a1**} \\ 
\underline{\beta} \ & = & \ O \ ( r^{- 2} \ \tau_-^{- \frac{3}{2}})  \label{b1**} \\ 
\rho 	\ & = & \ O \ ( r^{- 3})  \label{rhoA1}  \label{rho1**} \\ 
\rho - \bar{\rho} \ & = & \ O \ ( r^{- 3})  \label{rhoA2} \\ 
\sigma \ & = & \ O \ ( r^{- 3} \ \tau_-^{- \frac{1}{2}}) \label{sigmaA23}  \\ 
\sigma - \bar{\sigma}  \ & = & \ O \ ( r^{- 3} \ \tau_-^{- \frac{1}{2}}) \label{sigmaA2} \\ 
\beta \ & = & \ o \ (r^{- \frac{7}{2}})     \\ 
\alpha  \ & = & \ o \ (r^{- \frac{7}{2}})  
\eea 
for both (CK) and (A) spacetimes 
\bea  
\hat{\chi}  \ & = & \ O \ (r^{- 2})  \label{reshatchi1c} \\ 
\underline{\hat{\chi}}  \ & = & \ O \ (r^{-1} \tau_-^{- \frac{3}{2}}) \label{resuhatchi1c}  \\ 
\zeta, \ \epsilon, \ \delta  \ & = & \ O \ (r^{- 2})     \\ 
tr \chi \ & = & \  \frac{2}{r}  \ + \ l.o.t. \\ 
tr \underline{\chi} \ & = & \  - \frac{2}{r} \ + \ l.o.t.  \label{trchibar**1c}
\eea
for (CK) spacetimes 
$
\underline{\alpha} =  O ( r^{- 1} \tau_-^{- \frac{5}{2}})  , 
\underline{\beta} =  O ( r^{- 2} \tau_-^{- \frac{3}{2}}) , 
\rho = O ( r^{- 3})   , 
\rho - \bar{\rho} = O ( r^{- 3}  \tau_-^{- \frac{1}{2}})  , 
\sigma = O  ( r^{- 3}  \tau_-^{- \frac{1}{2}}) , 
\sigma - \bar{\sigma}  = O  ( r^{- 3}  \tau_-^{- \frac{1}{2}}) , 
\beta = o (r^{- \frac{7}{2}}) , 
\alpha =  o (r^{- \frac{7}{2}}) , 
$
for (B) spacetimes 
\beas
\underline{\alpha} \ & = & \ O \ ( r^{- 1} \ \tau_-^{- \frac{3}{2}})    \\ 
\underline{\beta} \ & = & \ O \ ( r^{- 2} \ \tau_-^{- \frac{1}{2}})    \\ 
\rho , \ \sigma , \ \alpha , \ \beta \ & = & \ o \ (r^{- \frac{5}{2}})     
\eeas 
and 
$
\hat{\chi}  = o (r^{- \frac{3}{2}})     , 
\underline{\hat{\chi}}  =  O (r^{-1} \tau_-^{- \frac{1}{2}})   , 
\zeta,  \epsilon,  \delta  = o (r^{- \frac{3}{2}})   , 
K  - \frac{1}{r^2}   = o (r^{- \frac{5}{2}})    , 
tr \chi  =  \frac{2}{r}   + l.o.t. , 
tr \underline{\chi}  =   - \frac{2}{r} +l.o.t.  , 
$
with $K$ the Gauss curvature of the surfaces $S_{t,u}$.

\subsection{Equations at Future Null Infinity $\mathcal{I}^+$}
\label{futureequs}

The shears are related to each other by the equation 
\be \label{chihat}
\frac{\partial}{\partial u} \hat{\chi} \ = \ \frac{1}{4} tr \chi \cdot \hat{\underline{\chi}} + l.o.t. 
\ee
Moreover, we have the equation 
\be \label{uchihat}
\frac{\partial}{\partial u} \underline{\hat{\chi}} \ = \ \frac{1}{2} \underline{\alpha} + l.o.t. 
\ee
Equations (\ref{chihat})-(\ref{uchihat}) are direct consequences from the structure equations and the faster fall-off behavior of the lower order terms involved. 

At this point, we recall the definitions for the Hodge duals. 
Let $W$ be a Weyl field on $M$, then define the left $\ ^*W$ and right $W^*$ Hodge duals as 
$
\ ^* W_{\alpha \beta \gamma \delta} :=  \frac{1}{2} \epsilon_{\alpha \beta \mu \nu} W^{\mu \nu}_{\ \ \ \gamma \delta} $ and $
W^*_{\alpha \beta \gamma \delta} := \frac{1}{2} W_{\alpha \beta}^{\ \ \ \mu \nu} \epsilon_{\mu \nu \gamma \delta} 
$
with $\epsilon^{\alpha \beta \gamma \delta}$ being the components of the volume element of $M$. 
Further, for a vectorfield $v$ on $S$ the Hodge dual $^*v$ is defined by 
$^*v_A := \epsilon_{AB} v^B$, where $\epsilon_{AB}$ denote the components of the area element relative to an arbitrary frame 
$\{ e_A \}$ with $A = 1,2$. 
For a symmetric, traceless $2$-tensor $\xi$ on $S$ the left $^*\xi$ and right $\xi^*$ Hodge duals are defined by 
$^*\xi_{AB} := \epsilon_{AB} \xi^C_B$, respectively $\xi^*_{AB} := \xi^C_A \epsilon_{CB}$.

Now, we recall the Bianchi identities \cite{ONeill}: 
\[
D_{[\epsilon} W_{\alpha \beta ] \gamma \delta} \ = \ 0 \ . 
\]
In the EV case, the Weyl curvature tensor satisfies (\ref{Bianchi2}) 
\[
D^{\alpha} W_{\alpha \beta \gamma \delta} \ = \ 0 \ . 
\]
Next, we are going to write these equations in the more suitable null frame, as introduced earlier. For our purpose, we need only two of these equations (\ref{Bianchiturho3})-(\ref{Bianchitusigma3}). For the full set of equations see \cite{sta}, \cite{lydia1}, \cite{lydia2}. The Bianchi identities for $\Dlap_3  \rho$, respectively $\Dlap_3 \sigma$ read 
\bea
&& \ \Dlap_3  \rho \ + \ \frac{3}{2} tr \underline{\chi} \rho \  =  \  \label{Bianchiturho3}  \\
& & \ 
 - \dlap \underline{\beta} 
 - \frac{1}{2} \hat{\chi} \underline{\alpha} \ + \ 
 ( \epsilon  - \zeta) \underline{\beta} \ + \ 
2 \underline{\xi} \beta 
 \nonumber \\ 
& & \ \Dlap_3 \sigma \ + \ \frac{3}{2} tr \underline{\chi} \sigma \  =  \  \label{Bianchitusigma3}  \\ 
& & \  - \clapa \ \ \underline{\beta}  - \frac{1}{2} \hat{\chi} ^*\underline{\alpha}  + 
\epsilon ^*\underline{\beta} - 2 \zeta ^*\underline{\beta} 
 - 2 \underline{\xi} ^*\beta  
 \nonumber 
\eea
where we define 
\[
 \underline{\xi}_A  =  \frac{1}{2} g(D_3 e_3, e_A) 
\]   
and $\epsilon_A = k_{AN}$.

In view of the electric-magnetic decomposition of the Weyl curvature tensor 
(\ref{Wel1})-(\ref{Wma1}), note that $\rho$ appearing in equation (\ref{Bianchiturho3}) is the following component of the electric part $E_{NN} = \rho$, 
and $\sigma$ appearing in equation (\ref{Bianchitusigma3}) is the following component of the magnetic part $H_{NN} = \sigma$. 
In the literature, memory involving the electric part of the Weyl curvature is typically called ``electric-parity memory", correspondingly for ``magnetic-parity memory". We shall derive the memory from equation (\ref{Bianchiturho3}) and show that the contribution from equation 
(\ref{Bianchitusigma3}) is zero for sources for which the metric falls off like mass/r towards infinity. Thus, the memory will be of purely electric parity for these sources.

\subsection{Future Null Infinity}
\label{future}

We recall that 
future null infinity $\mathcal{I}^+$ is defined to be the endpoints of all future-directed null geodesics along which $r \to \infty$. It has the topology of $\real \times \mathbb{S}^2$. Thus a null hypersurface $C_u$ intersects $\mathcal{I}^+$ at infinity in a $2$-sphere $S_{\infty, u}$. 
In this context, unless specified otherwise, the pointwise norms $| \ |$ of tensors on $S^2$ are with respect to the standard metric of $S^2$. We shall use $dS^2$ to denote the volume element for $S^2$ with respect to the standard metric. 
Correspondingly, in the following, all operators on $S^2$ such as $\dlap$, $\clapa \ $, $\nlap$, $\slap$ are automatically operators with respect to the standard metric on $S^2$. Christodoulou in \cite{chrmemory} denotes these with a superscript $^{0}$. However, in this article, we omit this notation for simplicity. But we keep in mind, that in this section, these operators applied to limits at future null infinity $\mathcal{I}^+$ are with respect to the standard metric on $S^2$.

It was established in \cite{sta}, \cite{chrmemory} and in \cite{lydia3} that for the scenarios investigated here, the following components attain limits at future null infinity $\mathcal{I}^+$, namely: 
The normalized curvature components $r\underline{\alpha }$, $r^{2}\underline{\beta } $, $r^3 \rho $, $r^3 \sigma $ have limits on $C_u$ as $t\rightarrow \infty $. 
Using (\ref{a1**})-(\ref{sigmaA2}) this yields: 
\begin{eqnarray*}
\lim_{C_{u},t\rightarrow \infty }r\underline{\alpha }_{AB}
&=& \underline{A}_{AB} ,\, \ \ \ \ 
\lim_{C_{u},t\rightarrow \infty }\,r^{2}\underline{\beta }_{A}
= \underline{B}_{A} \ , \\ 
\lim_{C_{u},t\rightarrow \infty} r^3 \rho & = & P \ ,  \ \ \ \ 
\lim_{C_{u},t\rightarrow \infty} r^3 \sigma  =  Q 
\end{eqnarray*}
where the limits are on $S^{2}$ and depend on $u$. These limits satisfy 
\begin{eqnarray*}
\left| A  \right| &\leq &C\left( 1+\left| u\right|
\right) ^{-5/2}\, \, \\
 \left| \underline{B} \right| & \leq & C\left( 1+\left| u\right| \right) ^{-3/2}   \ \ \\ 
\left| Q  \right| &\leq &C\left( 1+\left| u\right|
\right) ^{-1/2}\,
\end{eqnarray*}
whereas $P$ does not decay in $|u|$. 
For spacetimes investigated here, $(P (u, \theta_1, \theta_2) - \overline{P} (u))$ does not decay in $|u|$ either \cite{chrmemory}, \cite{lydia3}. 
Furthermore, the following limits exist and depend on the spherical variables as well as $u$. Using 
the results from \cite{chrmemory}, \cite{lydia3} yields the following limits. In particular using 
(\ref{reshatchi1c}) and (\ref{resuhatchi1c}) gives the following limits, that we denote by 
$\Sigma_{AB}$, respectively $\Xi_{AB}$: 
\bea 
\lim_{C_{u},t\rightarrow \infty } r^2 \widehat{\chi } & =: & \Sigma_{AB}  \label{limSigmaThe1}  \\ 
- \frac{1}{2} \lim_{C_{u},t\rightarrow \infty }r\widehat{\underline{\chi }} & =: & 
\Xi_{AB} \label{limXiThe1} 
\eea 
where 
\be \label{Xibsu}
\left| \Xi  \right| \leq C\left( 1+\left| u \right|
\right) ^{-3/2} 
\ee
Moreover, equation (\ref{1XiAT1}) follows from (\ref{uchihat}), equation (\ref{1SigmaT1}) from (\ref{chihat}), and from (\ref{codazzi2}) we obtain 
\begin{eqnarray}
\underline{B_{A}} &=& - 2 (\dlap \Xi)_{A} \label{Ldchibund1} \ . 
\end{eqnarray}

\subsection{Christodoulou Contribution to Memory}
\label{Christodouloumemory}

Let us use the notation $\rho_3 : = \Dlap_3  \rho \ + \ \frac{3}{2} tr \underline{\chi} \rho$ in (\ref{Bianchiturho3}) and 
$\sigma_3 : = \Dlap_3 \sigma + \frac{3}{2} tr \underline{\chi} \sigma$ in (\ref{Bianchitusigma3}). 

First, we consider the Bianchi equation for $\Dlap_3  \rho$, recalling (\ref{Bianchiturho3}). 
Looking at the right hand side, and using the information on how fast each term falls off towards infinity \cite{chrmemory}, \cite{lydia3}, 
we find that the leading order terms are the following (note that ``leading order terms" means terms of slowest fall-off towards infinity). 
\[
\rho_3 \  =  \ 
 - \underbrace{ \dlap \underline{\beta} }_{= O(r^{-3} \tau_-^{- \frac{3}{2}})}
 - \underbrace{ \frac{1}{2} \hat{\chi} \cdot \underline{\alpha} }_{= O(r^{- 3} \tau_-^{- \frac{5}{2}})} + \   O(r^{-4} \tau_-^{- \frac{3}{2}})  
\]
A short computation, using (\ref{chihat})-(\ref{uchihat}), shows that 
\beas
\rho_3 \  & = & \ 
 - \underbrace{ \dlap \underline{\beta} }_{= O(r^{-3} \tau_-^{- \frac{3}{2}})} 
- \underbrace{\frac{\partial}{\partial u} (\hat{\chi} \cdot \hat{\underline{\chi}})}_{ = O(r^{- 3} \tau_-^{- \frac{5}{2}})}  \\ 
& & + \underbrace{\frac{1}{4} tr \chi  |\hat{\underline{\chi}}|^2 }_{ = O(r^{-3} \tau_-^{-3})} + \   O(r^{-4} \tau_-^{- \frac{3}{2}})  
\eeas
Thus it is (omitting the terms of order $O(r^{-4} \tau_-^{- \frac{3}{2}}) $) 
\be \label{rho3*}
\rho_3    \  =  \ 
 - \dlap \underline{\beta} 
 + \frac{1}{4} tr \chi  |\hat{\underline{\chi}}|^2
- \frac{\partial}{\partial u} (\hat{\chi} \cdot \hat{\underline{\chi}})
 \ =  \ O(r^{-3} \tau_-^{- \frac{3}{2}}) 
\ee
Next, we take the limit of this equation at future null infinity $\mathcal{I}^+$. 
Thus, we multiply equation (\ref{rho3*}) by $r^3$ and take the limit along $C_u$ as $t \to \infty$ to obtain 
\be  \label{Brho3*2} 
- 2 \frac{\partial}{\partial u} P \  =  \ 
- \dlap \underline{B} + 2 \frac{\partial}{\partial u} ( \Sigma \cdot \Xi ) + 2 | \Xi |^2 
\ee
In the following, we denote by 
$\Sigma_{AB}^{+}$ the limit of $\Sigma_{AB}$ for $u \to + \infty$, 
respectively by $\Sigma_{AB}^{-}$
the limit of $\Sigma_{AB}$ for $u \to - \infty$. 
Correspondingly, we use the notation upper $^+$, respectively upper $^-$ for the corresponding limits of the quantities $P$, $Q$ and $\Xi$.

Then use (\ref{Ldchibund1}) as well as (\ref{1SigmaT1}) and integrate with respect to $u$ to get 
(for simplicity of notation omitting the angle-dependence) 
\bea
 (P (u) - P^+) \ & = & \ \dlap \dlap (\Sigma (u) - \Sigma^+)   \nonumber \\ 
& & + \int_{u}^{+ \infty} | \Xi (u') |^2 du' \nonumber \\ 
& &  - (  \Sigma (u) \cdot \Xi (u) -  \underbrace { \Sigma^+ \cdot \Xi^+}_{= 0}  )   \nonumber \\ 
\ &  = &  \ 
 \dlap \dlap (\Sigma (u) - \Sigma^+)    \nonumber \\ 
& & + \int_{u}^{+ \infty} | \Xi (u') |^2 du'     \nonumber \\ 
& & - \Sigma (u) \cdot \Xi (u)   \label{Pu1} \\ 
(P^- - P^+) \ &  = &  \ \dlap \dlap (\Sigma^- - \Sigma^+)   \nonumber \\ 
& & + \int_{- \infty}^{+ \infty} | \Xi (u) |^2 du  \nonumber \\ 
\ & & \  - ( \underbrace{ \Sigma^- \cdot \Xi^- }_{= 0}  - \underbrace{ \Sigma^+ \cdot \Xi^+ }_{= 0}  )  \nonumber \\ 
 \ & = & \ \dlap \dlap (\Sigma^- - \Sigma^+)  \nonumber \\ 
& & + \int_{- \infty}^{+ \infty} | \Xi (u) |^2 du   \label{Pall1}  
\eea
where also (\ref{Xibsu}) was used. 
Note that in (\ref{Pall1}) the last term is $2F$ from 
(\ref{recallF1}). 
Thus, using (\ref{recallF1}) we have 
\bea
\dlap \dlap (\Sigma^- - \Sigma^+) 
 \ & = & \ (P^- - P^+) 
- 2 F  \ .   \nonumber \\ \label{Pall2}
\eea 
As we said, $(\Sigma^- - \Sigma^+)$ gives the permanent displacement of the test masses in a detector. We see that there are two contributions: Namely, the null contribution from $-2F$ with $F$ given in (\ref{recallF1}), and the ordinary contribution from $(P^- - P^+)$ with the latter being the corresponding limits of the electric Weyl curvature component $\rho$. The former is Christodoulou's contribution to the memory, and the latter is the  piece of the memory that in the scenarios considered here and in our description using the full curvature tensor corresponds to the contribution found by Zel'dovich and Polnarev (recall that these authors studied a different situation and used an approximation method).

Next, we consider the 
Bianchi equation for $\Dlap_3 \sigma$ recalling (\ref{Bianchitusigma3}), 
Similarly as above, looking at the right hand side, and using the results from \cite{chrmemory}, \cite{lydia3}, (we also give the relevant information in the appendix (\ref{a1**})-(\ref{trchibar**1c})), we keep the leading order terms and write 
\[
\sigma_3 \ = \ - \clapa \ \ \underline{\beta} - \frac{1}{2} \hat{\chi} \cdot \ ^* \underline{\alpha} +  O(r^{-4} \tau_-^{- \frac{3}{2}})  
\]
A short computation yields (omitting the terms of order $O(r^{-4} \tau_-^{- \frac{3}{2}}) $) 
\be \label{sigma3*}
\sigma_3   \ = \ - \clapa \ \ \underline{\beta} -  \frac{\partial}{\partial u} ( \hat{\chi} \wedge \hat{\underline{\chi}} )\ = \ 
O(r^{-3} \tau_-^{- \frac{3}{2}})
\ee
Next, we take the limit of this equation at future null infinity $\mathcal{I}^+$. 
Thus, we multiply (\ref{sigma3*}) by $r^3$ and take the limit on $C_u$ as $r \to \infty$ to obtain 
\be 
Q_3 = - \clapa  \ \ \underline{B}  + 2 \frac{\partial}{\partial u} (\Sigma \wedge \Xi)
\ee
that reads 
\be \label{BSigmaauf*2*}
 \frac{\partial Q}{\partial u}  \ = \ \frac{1}{2}  \clapa  \ \ \underline{B}  -  \frac{\partial}{\partial u} (\Sigma \wedge \Xi) \ . 
\ee
Hence, it is 
\bea
\clapa \ \ \dlap (\Sigma^- - \Sigma^+)  \label{Qlimits*1*} \ & = & \   Q^- - Q^+  \ . \label{Qall2}
\eea
Summarizing, we have derived the system of equations on $S^2$ consisting of (\ref{Pall2}) and (\ref{Qall2}). 

The solution $(\Sigma^- - \Sigma^+)$ of these equations consists of the two pieces, that are the ordinary contribution (\ref{*P*}) and the null contribution (\ref{*C1*}) to the gravitational wave memory. 
Note that the first equation includes only the electric part $P$ of the Weyl tensor together with the energy $F$, whereas the second one features only the magnetic part $Q$ of the Weyl tensor. Therefore, equation (\ref{Pall2}) will give an electric parity memory, while (\ref{Qall2}) would give a magnetic parity memory but the latter will be zero for the scenarios considered here. 
In fact, using the results (\ref{sigmaA2}) and (\ref{sigmaA23}), it turns out that the right hand side of equation (\ref{Qall2}) is identically zero and therefore only equation (\ref{Pall2}) yields a memory. Thus, the memory is of electric parity only, while the magnetic parity part is zero. 
In the remainder of this section, in particular in the final formulas (\ref{*P*}) and (\ref{*C1*}), $\large< , \large>$ denotes the Euclidean inner product, and in the corresponding equations the norm $| \ |$ is taken with respect to the latter. 
We shall consider the scenarios where two bodies are coming in at non-relativistic speed, spiral around each other and finally merge to one body with final rest mass $M^*$ and with final (recoil) velocity $V$ relative to the initial center-of-mass frame. 

In order to solve the system ((\ref{Pall2}), (\ref{Qall2})) on $S^2$, first we derive a Hodge system from ((\ref{Pall2}), (\ref{Qall2})), and then we use Hodge theory to solve it. See \cite{chrmemory} or \cite{lydia3} for more details.

Finally, we are going to compute the solution 
$(\Sigma^- - \Sigma^+)(X, Y)$ of (\ref{Pall2}) and (\ref{Qall2}) at an arbitrary pair $X, Y$ of vectors in $\real^3$ tangent to $S^2$, at $\xi$, as it was done by Christodoulou in \cite{chrmemory}. Here, $\xi \in S^2 \subset \real^3$ is the direction of observation. 
Thus, the plane spanned by $(X, Y)$ is orthogonal to $\xi$. 
This solution is the sum of a contribution from 
\[
(P - P_{[1]})^- - (P - P_{[1]})^+ 
\]
and a contribution from 
\[
F - F_{[1]} 
\]
where subscript $[1]$ denotes the projection onto the sum of the $0^{th}$ $(l=0)$ and $1^{st}$ $(l=1)$ eigenspaces of $\slap$. 
This description is valid for all gravitational wave sources where the mass falls off towards infinity at a rate of ``mass/r". 

In the following, $\Pi$ is the projection to the plane orthogonal to $\xi$. 

In \cite{chrmemory} Christodoulou calculates the above contributions and the solution in the example of the binary coalescence to be as follows: 
\small{
\[
(\Sigma^- - \Sigma^+)(X, Y) = 
\]}
\[
\small{ - \frac{2 M^*}{( 1 - |V|^{2} )^{\frac{1}{2}}} \frac{<X, V> <Y, V> - \frac{1}{2} <X, Y> | \Pi V |^2}{1 - <\xi, V>} }
\]
\bea
\small{ - \frac{1}{2 \pi} \int_{|\xi'| = 1} 
\large(  F - F_{[1]} \large) (\xi') } \nonumber \\ 
\small{ \cdot \frac{\large< X, \xi' \large> \large< Y, \xi' \large> - \frac{1}{2} \large< X, Y \large> | \Pi \xi' |^2 }{1 - \large< \xi, \xi' \large>} 
d S^2 (\xi') } \nonumber \\ 
 \label{*C1*all} 
\eea

In (\ref{*C1*all}) the first term on the right hand side is the 
contribution from $(P - P_{[1]})^- - (P - P_{[1]})^+$: 
\be \label{*P*}
\small{ - \frac{2 M^*}{( 1 - |V|^{2} )^{\frac{1}{2}}} \frac{<X, V> <Y, V> - \frac{1}{2} <X, Y> | \Pi V |^2}{1 - <\xi, V>} }
\ee
and the second time on the right hand side is the 
contribution from $F - F_{[1]}$: 
\bea
\small{ - \frac{1}{2 \pi} \int_{|\xi'| = 1} 
\large(  F - F_{[1]} \large) (\xi') } \nonumber \\ 
\small{ \cdot \frac{\large< X, \xi' \large> \large< Y, \xi' \large> - \frac{1}{2} \large< X, Y \large> | \Pi \xi' |^2 }{1 - \large< \xi, \xi' \large>} 
d S^2 (\xi') } \nonumber \\ 
 \label{*C1*} 
\eea
Using $(\Sigma^- - \Sigma^+)(X, Y)$ in equation (\ref{pdispl*}), (\ref{*P*}) describes the ordinary contribution to memory and (\ref{*C1*}) describes the null contribution to memory. That is, the former corresponds to the contribution to memory derived by Zel'dovich and Polnarev \cite{zeldovichpolnarev} but applied to binary coalescence, the latter is the contribution to memory derived by Christodoulou \cite{chrmemory}. 

Now, let us consider the ordinary memory (\ref{*P*}) and the null memory (\ref{*C1*}) for different situations of gravitational wave bursts. 
\begin{itemize}
\item[(1)] Binary black hole mergers with recoil velocity $V \neq 0$: They have both contributions to memory (\ref{*P*}) and (\ref{*C1*}). 
\item[(2)] Binary black hole mergers without any recoil velocity, that is $V = 0$: Then the expression in (\ref{*P*}) is zero, whereas the expression in (\ref{*C1*}) is non-zero and in fact is large. 
\end{itemize} 
Moreover, note that for binary neutron star mergers, the contribution from (\ref{*P*}) may be small but the contribution from (\ref{*C1*}) is large. 
Equipped with these precise results, we revisit the discussion after formula (\ref{memfirst}) above.

\subsection{Types of Memory}
\label{typesm}

We have seen, that in the above scenarios, there is only electric-parity memory, but no magnetic-parity memory, as was found by Christodoulou \cite{chrmemory}. This follows directly from the results \cite{chrmemory}, \cite{sta}, and \cite{lydia3}.  
This is different from the situation for (B) spacetimes, as was shown more recently in 2020 by Bieri in \cite{lydia4}, \cite{lydia14}, where there is a magnetic memory, and both the electric and magnetic part grow with retarded time $\sqrt{| u |}$. In the latter papers, it was also concluded that data of slow decay where the initial metric behaves as follows for $r \to \infty$, namely $\bar{g}_{ij} - \eta_{ij} = o(r^{- \alpha})$ for 
$0 < \alpha < 1$, the two types of memory grow like $|u|^{1- \alpha}$. However, in this article, we focus on spacetimes for which the metric falls off like ``mass/r", as was considered by Christodoulou in \cite{chrmemory}.

\section{Hodge Theory}
\label{Hodge}

Here, we briefly sketch some well-known results from Hodge theory on the sphere $S^2$. These are used to solve the equations (\ref{Pall2}), (\ref{Qall2}). 

We first derive a Hodge system from equations (\ref{Pall2}), (\ref{Qall2}). Then we use Hodge theory to solve this.

Let $Z$ be a sufficiently smooth vector field on $S^2$. There exist scalar fields $\varphi$ and $\psi$ such that 
\[
Z = \nlap \varphi + \nlap^{\perp} \psi  \ . 
\]
Then the following equations hold 
\[
\dlap Z = \slap \varphi \ \ \ \ , \ \ \ \  \clapa \ \ Z = \slap \psi \ . 
\]
Next, we consider the equations on $S^2$ 
\bea
 \slap \varphi \ & = & \ f  \ , \label{f1} \\ 
 \slap \psi \ & = & \ g \ ,  \label{g1}
\eea
for sufficiently smooth functions $f, g$ with vanishing mean on $S^2$. 
By the Hodge theorem there exist smooth solutions to (\ref{f1}), respectively (\ref{g1}) that are unique up to an additive constant. 
In our article we consider the function $\varphi$ of vanishing mean of equation (\ref{f1}), whereas in section \ref{Christodouloumemory} the right hand side of equation (\ref{g1}) vanishes. \\

In section \ref{Christodouloumemory} we consider the situation for $Z$ given as 
\[
Z = \dlap (\Sigma^- - \Sigma^+) \ . 
\]

The above equations determine $(\Sigma^- - \Sigma^+)$ uniquely. \\

\section{Coupled Systems and Asymptotically-Flat Spacetimes of Slow Fall-off Towards Infinity}
\label{new1}

Many non-gravitational fields of physics that are coupled to the Einstein equations contribute to the null memory \cite{lbdg3}. This is true in particular for the Einstein-Maxwell system \cite{1lpst1}, \cite{1lpst2}, as well as for neutrino radiation \cite{lbdg1} as it occurs in a core-collapse supernova or a binary neutron star merger \cite{duez, st}, \cite{ott1}.

It was shown in \cite{lydia4, lydia14} 
that asymptotically-flat spacetimes with initial data metrics $\bar{g}_{ij}$ falling off towards infinity at a rate of $o(r^{- \frac{1}{2}})$ (rather than the rate in equation (\ref{imetric33})) and initial data $k_{ij}$ falling off towards infinity at a rate of $o(r^{- \frac{3}{2}})$ (rather than the rate in equation (\ref{iff33})) 
generate diverging magnetic-parity and electric-parity memory. In particular, magnetic-parity memory is not present for systems with metrics that fall off like $1/r$, but this shows prominently for systems with much slower fall-off. A scenario for this would be gravitational waves from a source (like a binary merger) within a large neutrino cloud.

\end{document}